\documentclass[aip,reprint]{revtex4-1}
\usepackage{graphicx}
\usepackage{amsmath, cases}
\usepackage{dcolumn}
\usepackage{bm}
\usepackage{ulem}
\usepackage{times}
\usepackage{amssymb}
\usepackage{epsfig}
\usepackage{xcolor}
\usepackage[T1]{fontenc}
\usepackage{mathtools, cases}   

\draft 

\begin{document}


\title{Tagged-particle motion of Percus-Yevick hard spheres from first principles} 



\author{Chengjie Luo}
\email[Electronic mail: ]{C.Luo@tue.nl}
\affiliation{ Soft Matter and Biological Physics, Department of Applied Physics, Eindhoven University of Technology, P.O. Box 513, 5600 MB Eindhoven, The Netherlands}
\author{Vincent E. Debets}
\email[Electronic mail: ]{V.E.Debets@tue.nl}
\affiliation{ Soft Matter and Biological Physics, Department of Applied Physics, Eindhoven University of Technology, P.O. Box 513, 5600 MB Eindhoven, The Netherlands}
\author{Liesbeth M.~C.~Janssen}
\email[Electronic mail: ]{L.M.C.Janssen@tue.nl}
\affiliation{ Soft Matter and Biological Physics, Department of Applied Physics, Eindhoven University of Technology, P.O. Box 513, 5600 MB Eindhoven, The Netherlands}


\date{\today}

\begin{abstract}
We develop a first-principles-based generalized mode-coupling theory (GMCT) for the tagged-particle motion of glassy systems. This theory establishes a hierarchy of coupled integro-differential equations for self-multi-point density correlation functions, which can formally be extended up to infinite order. We use our GMCT framework to calculate the self-nonergodicity parameters and the self-intermediate scattering function for the Percus-Yevick hard sphere system, based on the first few levels of the GMCT hierarchy. We also test the scaling laws in the $\alpha$- and $\beta$-relaxation regimes near the glass-transition singularity. Furthermore, we study the mean-square displacement and the Stoke-Einstein relation in the supercooled regime. 
We find that qualitatively our GMCT results share many similarities with the well-established predictions from standard mode-coupling theory, but the quantitative results change, and typically improve, by increasing the GMCT closure level. However, we also demonstrate on general theoretical grounds that the current GMCT framework is unable to account for violation of the Stokes-Einstein relation, underlining the need for further improvements in the first-principles description of glassy dynamics.

\end{abstract}

\pacs{}

\maketitle

\section{\label{sec:introduction} Introduction}
Glassy dynamics are displayed in many systems on different length scales such as atoms, colloids, granules, and even living cells \cite{berthier2011theoretical,janssen2019active}. Since the 1980s an increasing number of theories and models have been proposed to explain the emergence of glassy dynamics~\cite{Xia1999,Tarjus2005a,Sausset2008,Gotze1992,reichman2005mode,langer2014theories,biroli2013perspective}, but the nature of the glass transition is still fiercely debated today. One of the main challenges is to relate the tremendous slowdown of the dynamics of glass-forming materials to the only minor microstructural changes observed during supercooling or compression. The mode-coupling theory (MCT) of the glass transition is one of the few first-principles-based theories to describe glassy dynamics using only static structures as input \cite{leutheusser1984dynamical,bengtzelius1984dynamics}. This theory successfully qualitatively and semi-quantitatively predicts the intermediate scattering functions (ISF) and the self-intermediate scattering functions (SISF) from the static structure factors for many glass-forming materials, and as such is able to reproduce the celebrated scaling laws characterizing the two-step decay of the ISF and SISF near the glass transition point \cite{reichman2005mode,gotze2008complex,janssen2018mode}.

However, MCT invokes an uncontrolled factorization approximation which dismisses potentially important information covered in higher-order density correlations. Recently, a new theory called generalized mode-coupling theory (GMCT) has therefore been developed to remove or at least reduce the influence of this uncontrolled approximation \cite{szamel2003colloidal,wu2005high}. The basic idea is to develop exact equations of motion for the higher-order dynamic density correlations (instead of factorising them) so that a hierarchy of equations similar to MCT is developed. Early results have demonstrated that the predicted ISFs are systematically improved as more higher-order density correlators are included within the framework of GMCT \cite{janssen2015microscopic}. Moreover, the scaling laws predicted in MCT, which qualitatively agree with simulations and experiments, are preserved in GMCT with quantitatively improved exponent parameters \cite{luo2020generalized1,luo2020generalized2}.

Up until now, studies involving GMCT have only focused on the ISF, i.e.\ the dynamics of collective motion.\cite{szamel2003colloidal,wu2005high,mayer2006cooperativity,janssen2014relaxation,janssen2015microscopic,janssen2016schematic,luo2020generalized1,luo2020generalized2} The motion of a single tagged particle is, however, equally important and can reveal complementary information about the dynamics. From a practical point of view, the SISF can also offer better statistical quality. Indeed, since all particles are in principle statistically equivalent to the tagged one, it is generally much easier to reliably obtain the SISF from experiments or simulations than the ISF. This advantage of the SISF has rendered it a widely used quantity for comparison and testing of experiments and simulations with MCT during the past decades \cite{Walter1995Testing,weysser2010structural,Megen1998}. 

From both experiments and simulations it is known that the scaling laws of the SISF are similar to those of the ISF \cite{gotze2008complex}. In particular, there are two time scales for the two-step decay of the SISF: i) a time scale $\tau_\beta$ characterizing the $\beta$-relaxation regime, and ii) a time scale $\tau$ characterizing the $\alpha$-relaxation regime. Both $\tau_\beta$ and $\tau$ follow a power law with respect to the reduced packing fraction (or reduced temperature) and the exponents of these two power laws are related to each other. The dynamics of the SISF satisfy factorization scaling laws regarding wavenumbers and time in the $\beta$-relaxation regime and the time-density (or time-temperature) superposition principle in the $\alpha$-relaxation regime \cite{Walter1995Testing,gotze2008complex,Megen1998}. The dynamics of the $\alpha$-relaxation regime can also be well described by the stretched-exponential Kohlrausch function \cite{gotze2008complex}. All these scaling laws are successfully reproduced within MCT, but the exponent parameters are generally not accurate \cite{Walter1995Testing,voigtmann2004tagged,weysser2010structural}. To systematically improve upon these results a GMCT framework for the tagged-particle motion and the SISF is therefore required.

Another reason for developing an improved theory to predict the SISF is the need to provide better predictions regarding dynamical heterogeneity and in particular the Stokes-Einstein relation (SER). The original SER relates the single-particle diffusion coefficient $D$, the shear viscosity $\eta$, and the temperature $T$ to each other via $D\eta/T=\text{constant}$ \cite{Einstein1956}. Typically (although not always correctly), it is assumed that $\tau \propto \eta$ or $\tau \propto \eta/T$, which follows if the instantaneous shear modulus is temperature-independent or if a Gaussian solution to the diffusion equation is used, respectively \cite{shi2013relaxation}. In the present work we will assume these relations to hold and interpret $D\tau=\text{constant}$ 
as an effective SER. From simulations and experiments, however, it is known that the SER becomes significantly violated during vitrification \cite{tarjus1995breakdown,voigtmann2004tagged,shi2013relaxation,flenner2005relaxation,kumar2006nature}. This violation is generally regarded as a manifestation of dynamical heterogeneity, although some controversy still exists on its physical origins \cite{charbonneau2013dimensional}. 
Unfortunately, in MCT, where $\tau$ is obtained from the SISF and $D$ from the mean-squared displacement (MSD) (which is directly related to the SISF for vanishing wavenumber $k\rightarrow 0$), the violation of SER does not appear, or only shows up very weakly \cite{fuchs1998asymptotic,voigtmann2004tagged,flenner2005relaxation,weysser2010structural}. The relation between $D$ and $\tau$ within the supercooled regime thus requires further theoretical clarification.


The goal of this work is to develop and test GMCT for the tagged-particle motion. We take a monodisperse system of dense hard spheres as our reference system, with the static structure factors obtained analytically from the Percus-Yevick approximation \cite{wertheim1963exact}. We solve the SISF numerically within GMCT and comprehensively check the scaling laws of the SISF in the $\alpha$- and $\beta$-relaxation regimes for both liquid and glass states near the critical liquid-glass transition point. From the predicted SISF, we also calculate the MSD and carefully study the SER. Since a system of one-component hard spheres easily crystallizes at  high densities \cite{voigtmann2004tagged}, we compare our predictions to simulations of weakly polydisperse quasi-hard spheres \cite{weysser2010structural}. A study of multi-component GMCT will be published elsewhere \cite{ciarella2021multi}. 

This paper is organized as follows. We first briefly summarize the previously established GMCT framework for the collective particle motion, after which we present the formulation of GMCT for tagged-particle motion as well as the equations for the MSD. Next, we report the SISF predicted by GMCT for Percus-Yevick hard spheres, including the self-non-ergodicity parameters at the critical point, and the scaling laws for both liquid and glass states. We proceed by demonstrating the MSD and discussing the SER. Finally, we conclude our work with a critical evaluation.

\section{\label{sec:theory} Theory} 

\subsection{GMCT of collective motion}
We start by summarizing the microscopic GMCT equations of collective motion \cite{janssen2015microscopic}. Briefly, the theory seeks to describe spatiotemporal correlations in the density field (as encoded in the ISF) through the repeated and hierarchical application of the Zwanzig-Mori projection operator formalism \cite{zwanzig2001nonequilibrium,szamel2003colloidal}; each successive projector is built from an increasingly large basis of multi-point density modes. 
The dynamical objects of interest  are thus the $2n$-point density correlation functions
$F_n(\{k_i\}_{1\leq i \leq n},t)$, which are defined as 
\begin{equation}
\label{eq:Fndef}
F_n(\{k_i\}_{1\leq i \leq n}, t) = \langle \rho_{-\bm{k}_1}(0) \hdots \rho_{-\bm{k}_n}(0)
\rho_{\bm{k}_1}(t) \hdots \rho_{\bm{k}_n}(t) \rangle.
\end{equation} 
Here $\{k_i\}_{1\leq i \leq n}$ denotes a set of $n$ wavevectors $k_1,k_2,\hdots,k_n$, $\rho_{\bm{k}}(t)=\sum_{j=1}^{N_p} e^{i\bm{k}\cdot\bm{r}_j(t)}/\sqrt{N_p}$ represents a collective density mode with wavevector $\bm{k}$ at time $t$, $\bm{r}_j$ is the position of particle $j$, and $N_p$ is the total number of particles. The angular brackets denote an ensemble average, and the label $n$ ($n=1,\hdots,\infty$) specifies the level of the GMCT hierarchy. For convenience, we will neglect the subscript ${1\leq i \leq n}$ in the following. 
Note that for $n=1$, $F_1(k,t)$ is the usual intermediate scattering function.

In the overdamped limit, the GMCT equations read
\begin{gather} 
\nu_n\dot{F}_n(\{k_i\},t) + F_n(\{k_i\},t) S^{-1}_n(\{k_i\})J_n(\{k_i\})
\nonumber \\ 
+\int_0^t  \dot{F}_n(\{k_i\},t-u) J^{-1}_n(\{k_i\}) M_n(\{k_i\},u) du  = 0 \label{eq:GMCTF_n}, 
\end{gather} 
where $\nu_n$ is an effective friction coefficient, 
\begin{gather} 
\label{eq:Sn}
S_n(\{k_i\}) = F_n(\{k_i\},t=0)\approx \prod_{j=1}^n S(k_j)
\end{gather}
are $2n$-point static density correlation functions that serve as the $t=0$ boundary conditions and that are usually approximated in terms of the static structure factors $S(k_j)$,
and 
\begin{gather} 
\label{eq:Jn}
J_n(\{k_i\}) = \sum_{l=1}^n \frac{D_0k_l^2}{S(k_l)}\prod_{j=1}^n S(k_j)
\end{gather}
with $D_0$ denoting the bare diffusion coefficient. 
The memory functions are given by
\begin{gather}
M_n(\{k_i\},t) = \frac{\rho D_0^2}{2} 
\int \frac{d\bm{q}}{(2\pi)^3} \sum_{j=1}^n |V_{\bm{q,k}_j-\bm{q}}|^2
\hphantom{XXXX}
\nonumber \\
\times F_{n+1}(|\bm{k}_j-\bm{q}|,q,\{k_i\}^{(n-1)}_{i\neq j},t), \nonumber \\ \label{eq:Mn} 
\end{gather}
where $\rho$ is the number density, $\{k_i\}^{(n-1)}_{i\neq j}$ represents the set of $n-1$ wavenumbers in $\{k_1,\hdots,k_n\}$ except $k_j$, and $V_{\bm{q,k}_i-\bm{q}}$ are the static vertices that represent
wavevector-dependent coupling strengths.
The vertices are defined as
\begin{equation}
\label{eq:V}
V_{\bm{q,k-q}} = 
({\bm{k}} \cdot \bm{q}) c(q) + 
{\bm{k}} \cdot (\bm{k-q}) c(|\bm{k-q}|),
\end{equation}
with $c(q)$ denoting the direct
correlation function \cite{hansen1990theory}, which is related to the static structure factor via $c(q)
= [1-1/S(q)] / \rho$. Note that within GMCT the ISF is thus governed by $F_2(\{k_i\},t)$, which in turn is controlled by $F_3(\{k_i\},t)$, et cetera.
For additional details on e.g.\ the GMCT derivation, closure approximations, and numerical solutions for $F_n(\{k_i\},t)$, we refer to \cite{janssen2015microscopic,luo2020generalized1,luo2020generalized2}.

\subsection{GMCT of a tagged particle}
The dynamics of a tagged particle can be quantified by its SISF, $F^s_0(k_0,t)=\langle \rho^s_{-\bm{k}_0}(0)\rho^s_{\bm{k}_0}(t)\rangle$, where $\rho^s_{\bm{k}_0}(t)=e^{i\bm{k}_0\cdot\bm{r}(t)}$ is the tagged-particle density mode at wavevector $\bm{k}_0$ and $\bm{r}$ denotes the particle position. Note that we label the SISF with the subscript 0; generally we use the subscript $n$ to indicate the number of collective density modes in the correlator. 
The exact equation of motion for $F^s_0(k_0,t)$ is derived in MCT and given by \cite{fuchs1998asymptotic}
\begin{gather} 
\nu^s_0\dot{F}^{s}_0(k_0,t) + F^{s}_0(k_0,t) \left(S^{s}_0(k_0)\right)^{-1}J^s_0(k_0)
\nonumber \\ 
+\int_0^t  \dot{F}^s_0(k_0,t-u) \left(J^{s}_0(k_0)\right)^{-1} M^s_0(k_0,u) du  = 0 \label{eq:GMCTF_s0}.
\end{gather} 
Here $\nu^s_0$ is an effective friction coefficient for the tagged particle, $S^{s}_0(k_0)=1$, and $J^s_0(k_0)=D^s_0 k_0^2$ with $D^s_0$ the short-time diffusion coefficient of the tagged particle. Note that $D^s_0=D_0$ if we take the tagged particle to be one of the particles in the system. The memory function yields
\begin{gather}
\label{eq:Ms0} 
M^s_0(k_0,t) = \hphantom{XXXX}
\nonumber \\
\rho (D^s_0)^2 
\int \frac{d\bm{q}}{(2\pi)^3} |V^{s}_{\bm{q},\bm{k}_0}|^2
F^s_{1}(|\bm{k}_0-\bm{q}|,q,t),
\end{gather}
and is written in terms of the vertex
\begin{equation}
\label{eq:Vs}
V^{s}_{\bm{q},\bm{k}_0}=\bm{q}\cdot \bm{k}_0 c(q),
\end{equation}
and the tagged-particle four-point density correlation function
\begin{equation}
F^s_{1}(k_0,k_1,t)=\langle \rho^s_{-\bm{k}_0}(0)\rho_{-\bm{k}_1}(0) \rho^s_{\bm{k}_0}(t)
\rho_{\bm{k}_1}(t)  \rangle.
\end{equation}
Note that this density correlator contains both tagged-particle and collective density modes; physically, this is due to the fact that the motion of a single particle is also inherently governed by the collective dynamics of its environment.

In standard MCT, one uses the factorization approximation
\begin{equation}
F^s_{1}(k_0,k_1,t)\approx F^s_0(k_0,t)\times F_1(k_1,t),
\end{equation}
such that Eq.\ (\ref{eq:GMCTF_s0}) for $F^s_0(k_0,t)$ can be solved self-consistently together with the MCT equation for $F_1(k_1,t)$. However, the factorization approximation is uncontrolled, and here we avoid it by instead developing an exact equation of motion for $F^s_{1}(k_0,k_1,t)$. 
Generalizing this concept, we can define the $n$th order tagged-particle multi-point density correlation functions, which read
\begin{gather}
F^{s}_n(k_0,\{k_i\},t)=
\nonumber \\
\langle \rho^s_{-\bm{k}_0}(0)\rho_{-\bm{k}_1}(0) \hdots \rho_{-\bm{k}_n}(0)\rho^s_{\bm{k}_0}(t)
\rho_{\bm{k}_1}(t) \hdots \rho_{\bm{k}_n}(t) \rangle, \ n\geq 1.
\end{gather}
Here our notation convention for $n$ implies that the correlators $F_n(\{k_i\},t)$ and $F^{s}_n(k_0,\{k_i\},t)$ contain the same number of collective density modes; the difference between $F_n(\{k_i\},t)$ and $F^{s}_n(k_0,\{k_i\},t)$ is thus $\rho^s_{-\bm{k}_0}(0) \rho^s_{\bm{k}_0}(t)$.
The exact equations of motion for $F^{s}_n(k_0,\{k_i\},t)$ can be derived using the Zwanzig-Mori projection operator formalism analogous to the collective case \cite{janssen2015microscopic}, and they take on a similar form, i.e.\ 
\begin{gather} 
\nu^s_n\dot{F}^{s}_n(k_0,\{k_i\},t) 
\nonumber \\
+ F^{s}_n(k_0,\{k_i\},t) \left(S^{s}_n(k_0,\{k_i\})\right)^{-1}J^s_n(k_0,\{k_i\})
\nonumber \\ 
+\int_0^t  \dot{F}^s_n(k_0,\{k_i\},t-u) \left(J^{s}_n(k_0,\{k_i\})\right)^{-1} M^s_n(k_0,\{k_i\},u) du 
\nonumber \\
 = 0 \label{eq:GMCTF_sn},
\end{gather} 
where
\begin{equation}
S_n^s(k_0,\{k_i\})=F^{s}_n(k_0,\{k_i\},0)=S_n(\{k_i\}) \approx \prod_{j=1}^n S(k_j), n \geq 1,
\end{equation}
and 
\begin{equation}
J_n^s(k_0,\{k_i\})=D^s_0 k_0^2 \prod_{j=1}^n S(k_j) +\sum_{l=1}^n \frac{D_0k_l^2}{S(k_l)} \prod_{j=1}^n S(k_j), n \geq 1.
\end{equation}
Notably, the memory functions for the higher order tagged-particle density correlators consist of two distinct contributions (in contrast to e.g.\ the collective memory function $M_n$) which are given by 
\begin{gather}
M^s_n(k_0,\{k_i\},t) = \hphantom{XXXX}
\nonumber \\
\rho (D^s_0)^2 
\int \frac{d\bm{q}}{(2\pi)^3} |V^{s}_{\bm{q},\bm{k}_0}|^2
F^s_{n+1}(|\bm{k}_0-\bm{q}|,q,\{k_i\},t) + \nonumber \\
\frac{\rho D_0^2 }{2}
\int \frac{d\bm{q}}{(2\pi)^3} \sum_{j=1}^n |V_{\bm{q,k}_j-\bm{q}}|^2
F^s_{n+1}(k_0,|\bm{k}_j-\bm{q}|,q,\{k_i\}^{(n-1)}_{i\neq j},t).\label{eq:Msn} 
\end{gather}
Here $V_{\bm{q,k}_j-\bm{q}}$ is the same vertex  as in the collective case [Eq.~(\ref{eq:V})] and  $V^{s}_{\bm{q},\bm{k}_0}$ is the self vertex, Eq.~(\ref{eq:Vs}), in the memory function of MCT for the tagged particle.

In principle, the hierarchy of GMCT equations can continue up to infinite order, but in practice, due to limited computing power, we have to choose a closure at a finite level $N$, similar to what has been done in GMCT for the collective particle motion \cite{janssen2015microscopic}. Following earlier work, we may consider two types of  closure approximations.\cite{janssen2016schematic,biezemans2020glassy} One type is the exponential closure, which simply cuts off the hierarchy by setting $F^s_N(k_0,\{k_i\},t)=0$, such that $F^s_{N-1}\sim \exp(t/\tau_{N-1})$ with $\tau_{N-1}=\nu^s_{N-1}/D^s_0k_0^2$. As a consequence, ${F}^{s}_n(t)$ will always decay to zero, and this GMCT closure approximation typically leads to an underestimation of the true dynamics, i.e.\ yielding too fast relaxation. The other type of closure is a so-called mean-field (MF) closure, approximating $F^s_N(k_0,\{k_i\},t)$ by (a combination of) density correlation functions at lower levels. An example of such a MF closure is 
\begin{equation}
F^s_N(k_0,\{k_i\},t)=F^s_0(k_0,t)F_N(\{k_i\},t).
\label{eq:Fsclosure}
\end{equation}
In contrast to the exponential closure, an MF closure can produce a sharp glass transition but it tends to overestimate the dynamics, thus generally predicting too slow relaxation. When $N\rightarrow\infty$ we expect the GMCT predictions under the two types of closures to converge \cite{janssen2015microscopic,janssen2016schematic}. In this work we will only report the dynamics using MF closures of the form (\ref{eq:Fsclosure}), for which a liquid-glass transition can indeed be predicted.

\subsection{Mean-squared displacement}
The equation of motion for the MSD $\delta r^2(t)=\langle|\bm{r}(t)-\bm{r}(0)|^2\rangle$ can be derived from Eq.~(\ref{eq:GMCTF_s0}) by expanding the tagged-particle density correlator up to second order, i.e.\  $F^s_0(k_0,t)=1-k_0^2\delta r^2(t)/6+\mathcal{O}(k_0^4)$, and taking the limit $k_0\rightarrow 0$. If we assume our particles to be Brownian ($\nu_0^s=1$), this results in
\begin{equation}
\delta r^2(t)+D_0^s \int_0^t M_{\text{MSD}}(t-u)\delta r^2(u)du=6D_0^s t,
\label{eq:MSD}
\end{equation}
where
\begin{gather}
\label{eq:Mmsd_middle} 
M_{\text{MSD}}(t)= \lim_{k_0\rightarrow 0}M^s_0(k_0,t)/(D_0^s)^2k_0^2 = \hphantom{XXXX}
\nonumber \\
\lim_{k_0\rightarrow 0} \rho 
\int \frac{d\bm{q}}{(2\pi)^3} \frac{|V^{s}_{\bm{q},\bm{k}_0}|^2}{k_0^2}
F^s_{1}(|\bm{k}_0-\bm{q}|,q,t).
\end{gather}
Carrying out the limit $k_0\rightarrow 0$ we finally obtain the memory functions for the MSD 
\begin{gather}
\label{eq:Mmsd} 
M_{\text{MSD}}(t)= \frac{\rho }{6\pi^2}
\int_0^{\infty}  p^4 c^2(p)
F^s_{1}(p,p,t) dp.
\end{gather}
Using Eqs.\ (\ref{eq:MSD}) and (\ref{eq:Mmsd}) with the calculated  $F^s_{1}(p,p,t)$ from GMCT for the tagged-particle motion as an input, the MSD can be calculated numerically.

There is a diffusion-localization transition for the MSD, which can be characterized by its long-time limit. In the liquid state, the particle always migrates diffusively in the long-time limit with a long-time tagged-particle diffusion coefficient $D^s$, i.e.
\begin{equation}
\lim_{t\rightarrow\infty}\delta r^2(t)=6D^s t,
\label{eq:Ds}
\end{equation}
where 
\begin{equation}
D^s =D_0^s \left[1+D_0^s\int_{0}^{\infty}M_{MSD}(u)du\right]^{-1}.
\label{eq:Ds_M}
\end{equation}
However, in the glass state the tagged particle is localized, which is manifested by a constant MSD at long times. A characteristic localization length $r_s$ can then be defined via \cite{fuchs1998asymptotic}
\begin{equation}
F^s_0(k_0)=1-(k_0r_s)^2+\mathcal{O}(k_0^4),
\end{equation}
which corresponds to the following asymptotic value of the MSD 
\begin{equation}
\lim_{t\rightarrow\infty}\delta r^2(t)=6r_s^2.
\label{eq:rs}
\end{equation}
Invoking Eq.~(\ref{eq:MSD}) we obtain the following expression for $r_s$ 
\begin{equation}
r_s^2=1/M_{MSD}(t\rightarrow\infty).
\end{equation}

\subsection{Numerical details}
Overall, Eqs.\ (\ref{eq:GMCTF_n}), (\ref{eq:GMCTF_s0}), (\ref{eq:GMCTF_sn}), and (\ref{eq:MSD}) allow us to obtain the ISF, SISF, and MSD for a glass-forming material using the corresponding static structure factors $S(k)$ as the only input. Here we study these GMCT equations for a monodisperse Percus-Yevick hard sphere system using the packing fraction $\varphi$ as our control parameter. We numerically solve the dynamical equations on an equidistant wavenumber grid of 100 points ranging from $kd=0.2$ to $kd=39.8$, where $d$ is the diameter of the particle. The integrals over the wavevector $\bm{q}$ in the memory functions Eqs.~(\ref{eq:Mn}) and (\ref{eq:Msn}) and $p$ in Eq.~(\ref{eq:Mmsd}) are approximated as a double and single Riemann sum, respectively \cite{franosch1997asymptotic}. For the time-dependent integration, we apply the fast algorithm of \cite{flenner2005relaxation,fuchs1991comments} with a starting time step size $\Delta t=10^{-6}$ that is doubled every $32$ points. We assume $D_0=D_0^s=1$ and set $\nu_n=\nu_n^s=1$ for all $n$. 
%
%

\section {\label{sec:results} Results and discussion}
\subsection{ Self-non-ergodicity parameters}
\begin{figure}[h!]
	\includegraphics{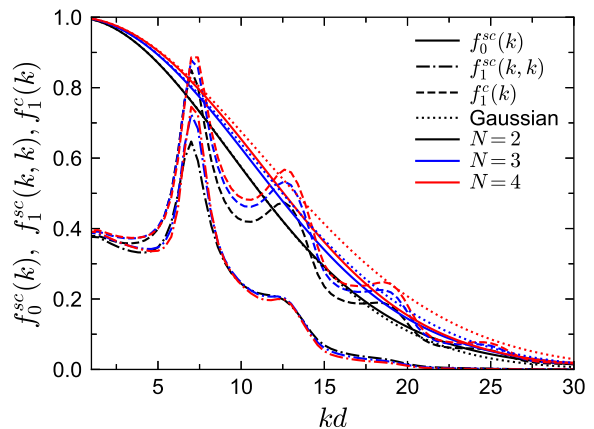}
	\caption{\label{fig:fc} The collective non-ergodicity parameters $f_1^c(k)$ and self-non-ergodicity parameters $f^{sc}_0(k)$, $f^{sc}_1(k,k)$ for Percus-Yevick hard spheres as a function of wavenumber $k$ at the critical packing fractions $\varphi^c$ for different GMCT MF closure levels $N$.}
\end{figure}

\begingroup
\setlength{\tabcolsep}{8pt} 
\renewcommand{\arraystretch}{1.1} %
\begin{table}
	\caption{Predicted critical packing fractions $\varphi^c$ and parameters $\gamma$, $a$, $b$, $\lambda$ determined from the SISFs for Percus-Yevick hard spheres obtained 
		for different GMCT mean-field closure levels $N$. The data for $N=5$ are taken from \cite{luo2020generalized1} for the ISF and are assumed to be preserved for the SISF.
	}
	\begin{tabular}{ccccccc} 
		\hline
		\hline
		\\[-1em]
		$N$ & $\varphi^c$&$\gamma$&$a$&$b$&$\lambda$&$r_s$\\
		\\[-1em]
		\hline
		\\[-1em]
		$2$ &0.515914 &2.46&0.31&0.59&0.73&0.0744\\
		$3$ & 0.533862 &2.71&0.29&0.51&0.78& 0.0667\\
		$4$ & 0.546851 &2.95&0.27&0.45&0.81& 0.0627\\
		$5$  & 0.556824 &3.15&0.25&0.43&0.83&\\
		\\[-1em]
		\hline
		\hline
		
	\end{tabular}
	
	\label{tab:1}
\end{table}
\endgroup

We first consider the long-time limit of the SISF, i.e.\ the self-non-ergodicity parameter, at the glass transition point.  We define the critical point $\varphi^c$ as the lowest packing fraction where the long-time limit of the SISF is larger than zero. From previous GMCT studies for the collective dynamics of hard spheres \cite{luo2020generalized1,szamel2003colloidal,wu2005high}, it is known that the critical glass transition point increases with the GMCT closure level $N$ in a seemingly convergent manner. Here we find identical values for $\varphi^c$ as predicted from the SISF, see Table \ref{tab:1}. We note, however, that in general the location of the glass transition point can be different for the ISF and SISF, as is the case for e.g.\ binary hard-sphere mixtures \cite{voigtmann2011multiple}.  
In Fig.~\ref{fig:fc} we plot the self-non-ergodicity parameters $f^{sc}_0(k)$ (solid lines) at the respective GMCT critical points for $N=2,\ 3,\ 4$. For comparison we also show the corresponding collective non-ergodicity parameters $f_1^c(k)$ (dashes lines in Fig.~\ref{fig:fc}). It can be seen that $f^{sc}_0(k)$ increases with increasing $N$ for nearly all wavenumbers $k$, which physically indicates relatively slower dynamics of a tagged particle. This trend is also consistent with the behavior seen in the collective density correlators $f_1^c(k)$ \cite{luo2020generalized1}.


In simulations it can be difficult to retrieve $f^{sc}_0(k)$ precisely, and instead one often uses the plateau values $A(k)$ obtained from fitting the SISF to the so-called Kohlrausch and von Schweidler functions [see also Eq.~(\ref{eq:Kohlrausch})]. These plateau values should be roughly equal to $f^{sc}_0(k)$ albeit slightly smaller \cite{fuchs1994kohlrausch,weysser2010structural}. However, previous simulations of weakly polydisperse hard spheres \cite{weysser2010structural} have demonstrated that, for wavenumbers $kd>10$, the theoretically predicted $f^{sc}(k)$ from MCT is in fact smaller than the corresponding $A(k)$. This discrepancy could be fixed using GMCT since the self-non-ergodicity parameters $f^{sc}(k)$ predicted from higher-order GMCT are larger than the ones from MCT, and they can also become larger than the $A(k)$ from simulations. This needs to be tested with a GMCT for multicomponent systems in the future \cite{ciarella2021multi}, but the trend of our $f^{sc}(k)$ suggests that GMCT could (at least partly) remedy this inconsistency. 

In MCT, i.e.\ when $N=2$, the $f^{sc}_0(k)$ can be well described by a Gaussian $f^{sc}_0(k)\approx e^{-k^2r_{sc}^2}$ up to $kd=15$ (black dotted lines in Fig.~\ref{fig:fc}) \cite{fuchs1998asymptotic}. However, with higher closure level $N$, deviations from the Gaussian approximation start to occur at much smaller wavenumbers and also become larger (blue and red dotted lines in Fig.~\ref{fig:fc}). This might reflect some traits of  dynamical heterogeneity, i.e.\ non-trivial spatiotemporal correlations in the density, which could be embedded in the higher-order density correlators that are explicitly included in GMCT. 

Finally, we have also plotted the long-time limit of the density correlators one level above $f^{sc}_0(k)$, i.e.\ $f_1^{sc}(k,k)$, in Fig.~\ref{fig:fc} (dash-dotted lines). At all closure levels $N$, $f_1^{sc}(k,k)$ is modulated by the structure of $S(k)$ with a maximum at $kd\approx7.4$. This is natural for MCT since $f_1^{sc}(k,k)=f_0^{sc}(k)\times f^c_1(k)$ and $f^c_1(k)$ is strongly modulated by $S(k)$. A salient feature of the $f_1^{sc}(k,k)$ GMCT curves in Fig.~\ref{fig:fc} is that they do not show a monotonic change with the closure level $N$; rather, we find crossovers at several wavenumbers $k$ where the ordering of the $f_1^{sc}(k,k)$ curves changes. We speculate that this behavior, which is also exhibited by the collective GMCT analogues $f^c_2(k,k)$ \cite{luo2020generalized2}, might hint at the presence of dynamical heterogeneities, but further research is needed to elucidate the microscopic origins of these $N$-dependent crossovers.

\subsection{ Time-dependent relaxation dynamics of SISF}
\begin{figure}[h!]
	\includegraphics{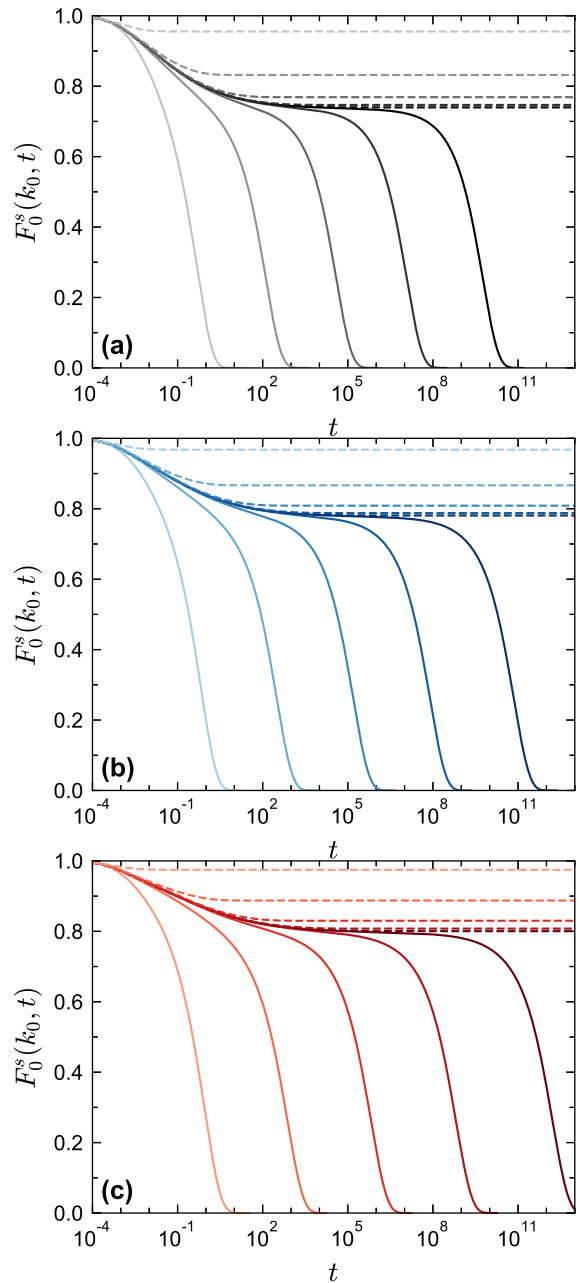}
	\caption{\label{fig:dynamics} The self-intermediate scattering function $F_0^s(k_0,t)$ at $k_0d=7.4$ for Percus-Yevick hard spheres obtained from  GMCT under MF closure levels (a) $N=2$, (b) $N=3$, (c) $N=4$. In all panels, the SISF is plotted at $|\epsilon|=10^{-1},\ 10^{-2},\ 10^{-3},\ 10^{-4},\ 10^{-5}$ (from light to dark) for both liquid states (solid lines) and glass states (dashed lines).}
\end{figure}

We now seek to study the time-dependent self-intermediate scattering functions $F_0^s(k_0,t)$, which have been obtained by solving Eqs.~(\ref{eq:GMCTF_s0}) and (\ref{eq:GMCTF_sn}) with the collective multi-point density correlators $F_N(\{k_i\},t)$ solved from Eq.~(\ref{eq:GMCTF_n}) as inputs. Figure \ref{fig:dynamics} shows the Percus-Yevick hard-sphere $F_0^s(k_0,t)$ for several reduced packing fractions $\epsilon=(\varphi-\varphi^c)/\varphi^c$ under different MF closure levels $N$. For liquid states, i.e.\ when the packing fraction is below the corresponding GMCT critical point,  $F_0^s(k_0,t)$ decays from $1$ to $0$ (solid lines). However, for glass states with $\varphi \geq \varphi^c$,  $F_0^s(k_0,t)$ decays from $1$ to the positive long-time limit $f_0^s(k_0)$ (dashed lines). In the following we will numerically show the scaling laws for these two states within GMCT.

\subsubsection{Relaxation dynamics for liquid states}
Let us first introduce some common properties of the liquid-state relaxation dynamics for all MF closure levels within GMCT. When the packing fraction is small and we are far from the critical point, the relaxation is governed by a relatively fast exponential decay. In contrast, when the packing fraction approaches the critical point, $F_0^s(k_0,t)$ exhibits a two-step decay: i) an initial decay from $1$ to a plateau very close to the self-non-ergodicity parameter $f_0^{sc}(k_0)$ characterized by a time scale $\tau_{\beta}$; and ii) a final decay from the end of the plateau to $0$ with a typical time scale $\tau$. The regime near the plateau is called the $\beta$-relaxation regime and the regime of the second decay is called the $\alpha$-relaxation regime. 

\begin{figure}[h!]
	\includegraphics{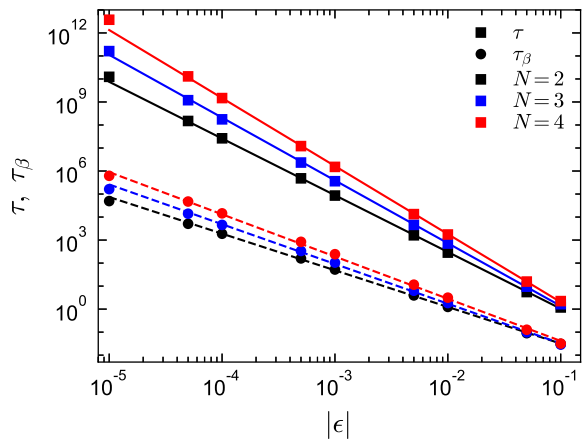}
	\caption{\label{fig:tau_scaling} Relaxation times of the self-intermediate scattering functions $F^s_0(k,t)$ at $kd = 7.4$ for different GMCT MF closure levels. Squares are the numerical $\alpha$-relaxation times and circles are the numerical $\beta$-relaxation times. The solid lines are the fitted power-law curves $\tau=\tau_0 |\epsilon|^{-\gamma}$. The dashed lines are the fitted power-law curves $\tau_{\beta}=\tau_{\beta 0}|\epsilon|^{-1/2a}$. The $\tau_0$, $\tau_{\beta 0}$, $\gamma$ and $a$ are all $N$-dependent. }
\end{figure}

From the relaxation curves in Fig.~\ref{fig:dynamics} we can extract the $\beta$-relaxation time scale $\tau_{\beta}$ (circles in Fig.~\ref{fig:tau_scaling}) defined as $F_0^s(k_0,\tau_{\beta})=f_0^{sc}(k_0)$. In standard MCT, $\tau_{\beta}$ can be well described by a power law $\tau_{\beta}\sim \epsilon^{-1/2a}$ with $a$ an exponent determined solely by the critical point $\varphi^c$. We find this power law to also be applicable for GMCT with closure levels $N\geq 2$. However, the exponent $a$ now depends on both $N$ and the corresponding critical point, similar to what has been reported recently for the collective GMCT dynamics \cite{luo2020generalized1,luo2020generalized2}. We observe that as $N$ increases, $a$ decreases, indicating relatively slower tagged-particle dynamics with respect to the critical point.

The $\alpha$-relaxation time scale $\tau$, which we define as $F_0^s(k_0,\tau)=0.1$, can also be obtained from Fig.~\ref{fig:dynamics}  as a function of $\epsilon$ (squares in Fig.~\ref{fig:tau_scaling}). The resulting values also obey a power law relation, i.e.\ $\tau\sim \epsilon^{-\gamma}$, for all MF closure levels $N$. Similar to the exponent $a$, $\gamma$ depends on $N$, although its value instead increases for larger $N$. This corresponds to a lower fragility, as has already been pointed out in our previous study \cite{luo2020generalized1}. 

The two exponents $a$ and $\gamma$ are related to each other via \cite{gotze2008complex,fuchs1998asymptotic} 
\begin{equation}
\lambda=\frac{\Gamma(1-a)^2}{\Gamma(1-2a)}=\frac{\Gamma(1+b)^2}{\Gamma(1+2b)},
\end{equation}
and
\begin{equation}
\gamma=\frac{1}{2a}+\frac{1}{2b},
\end{equation}
with $0<\lambda<1$, $0<a<1/2$, and $b>0$. Based on the obtained values for $a$ and $\gamma$, the additional parameters have also been retrieved, see Table \ref{tab:1}. Note that in principle one only needs one of the exponents to calculate  the others.
Importantly, we find that for all considered values of $N$, all parameters are identical to the ones from the GMCT dynamics of the collective density correlators \cite{luo2020generalized1}. (This can also be strictly proved by expanding the SISF close to the critical point, see  \cite{luo2020generalized2} for more details). Thus, higher-order GMCT seems to maintain the equivalency of the critical exponents $a$, $b$, $\gamma$, and $\lambda$ between collective and tagged-particle motion, a result that is also obtained in standard MCT.

\begin{figure}[h!]
	\includegraphics{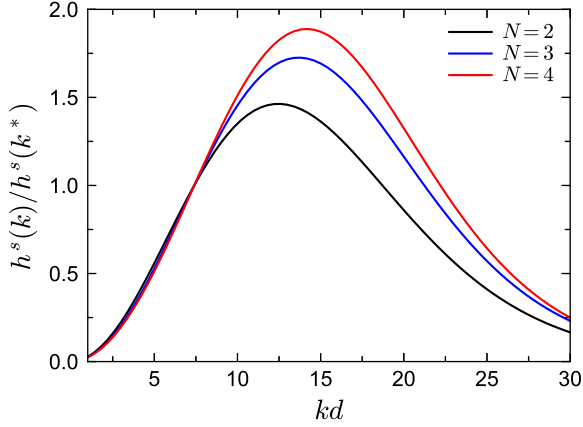}
	\caption{\label{fig:hsq} The amplitude $h^s(k)/h^s(k^*)$ with $k^*d=7.4$ for different GMCT MF closure levels $N$.}
\end{figure}

Now let us focus on the $\beta$-relaxation regime in more detail. Within MCT, to leading order in $|\epsilon|$, there is a scaling law for $F_0^s(k_0,t)$ near the plateau, i.e.
\begin{equation}
F_0^s(k,t)= f_0^{sc}(k)+h^s(k) G(t).
\label{eq:beta_scaling}
\end{equation}
Here $G(t)$ is the so-called $\beta$-correlator (note that it is independent of the wavenumber $k$), which satisfies
\begin{equation}
G(t)\sim \sqrt{|\epsilon|} g_-(t/\tau_{\beta}).
\label{eq:beta_scaling_epsilon}
\end{equation}
and, when $\epsilon\approx 0-$, 
\begin{equation}
G(t)\sim \left\{
\begin{aligned}
&t^{-a}& \textnormal{if} \ t<\tau_{\beta}, \\
&t^{b}&\textnormal{if} \ t>\tau_{\beta}.
\end{aligned}
\right.
\label{eq:beta_scaling_power}
\end{equation}
We find that these scaling laws are preserved within GMCT for all MF closure levels $N\geq 2$, although the factors $f_0^{sc}(k)$, $h^s(k)$, and the function $g_-(t/\tau_{\beta})$ become $N$-dependent. 
In order to demonstrate the respective scaling relations we have first retrieved the critical amplitude $h^s(k)$ for different $N$. This has been done by simply taking two times $t_1$ and $t_2$ within the $\beta$-relaxation regime and using
\begin{equation}
\frac{h^s(k)}{h^s(k^*)}=\frac{F_0^s(k,t_1)-F_0^s(k,t_2)}{F_0^s(k^*,t_1)-F_0^s(k^*,t_2)},
\end{equation}
which can be derived from Eq.~(\ref{eq:beta_scaling}).
Figure \ref{fig:hsq} shows ${h^s(k)}/{h^s(k^*)}$ with $k^*d=7.4$ for $N=2,3$, and $4$.  
It can be seen that both the height of the peak of the scaled amplitude $h^s(k)/h^s(k^*)$ and the wavenumber corresponding to the peak increase as $N$ increases. We know from the weakly polydisperse hard-sphere simulations of \onlinecite{weysser2010structural} that the MCT-predicted wavenumber of the peak of $h^s(k)/h^s(k^*d=7.4)$ is already larger than the one from simulation data (see Fig.~11 in \onlinecite{weysser2010structural}). The value of the peak is also overestimated by MCT. Therefore, for increasing closure level $N$, more pronounced deviations from the simulation data seem to occur, and our current GMCT framework is manifestly unable to reach improved agreement. We can explain the shift in the peak wavenumber by realizing that the $\beta$-relaxation regime corresponds to the caging effect \cite{janssen2018mode,kob2002supercooled} and $h^s(k)$ measures the decay of the SISF within this regime. We expect that the strongest decay occurs on length scales of the cage size, which would imply that higher-order GMCT predicts a smaller cage size. This will be confirmed when we study the MSD (see Fig.\ \ref{fig:msd_critical}). Moreover, the fact that the peak shifts to larger wavenumbers also means that $h^s(k^*d=7.4)$ takes on a relatively smaller value, which can explain why the height of the peak increases.   

\begin{figure}[h!]
	\includegraphics{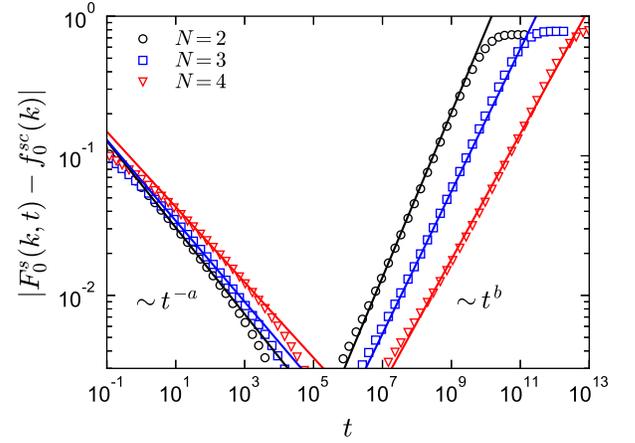}
	\caption{\label{fig:critical_dynamics} Relative self-intermediate scattering functions $|F_0^s(k,t)-f^{sc}_0(k)|$ at $kd=7.4$ and $\epsilon=-10^{-5}\approx 0^{-}$. The symbols represent the numerical GMCT critical dynamics for $|F_0^s(k,t)-f^{sc}_0(k)|$. The solid and dashed lines are fits of $|F_0^s(k,t)-f^{sc}_0(k)|\sim t^{-a}$ and $|F_0^s(k,t)-f^{sc}_0(k)|\sim t^{b}$, respectively.}
\end{figure}

The scaling laws of Eqs.\ (\ref{eq:beta_scaling})--(\ref{eq:beta_scaling_power}) in the $\beta$-relaxation regime can also be tested numerically from our GMCT SISF results for Percus-Yevick hard spheres. To test the power-law decay of Eq.~(\ref{eq:beta_scaling_power}), we set $\epsilon=-10^{-5}\sim 0^-$ to ensure a sufficiently close proximity to the critical point.  Figure \ref{fig:critical_dynamics} shows the relative self-intermediate scattering functions $|F_0^s(k,t)-f^{sc}_0(k)|$ at wavenumber $kd=7.4$ for different closure levels $N$. All curves conform to power laws with exponents $-a$ and $b$  at the beginning and end of the $\beta$-relaxation regime, respectively, thus confirming the validity of Eq.~(\ref{eq:beta_scaling_power}).

\begin{figure}[h!]
	\includegraphics{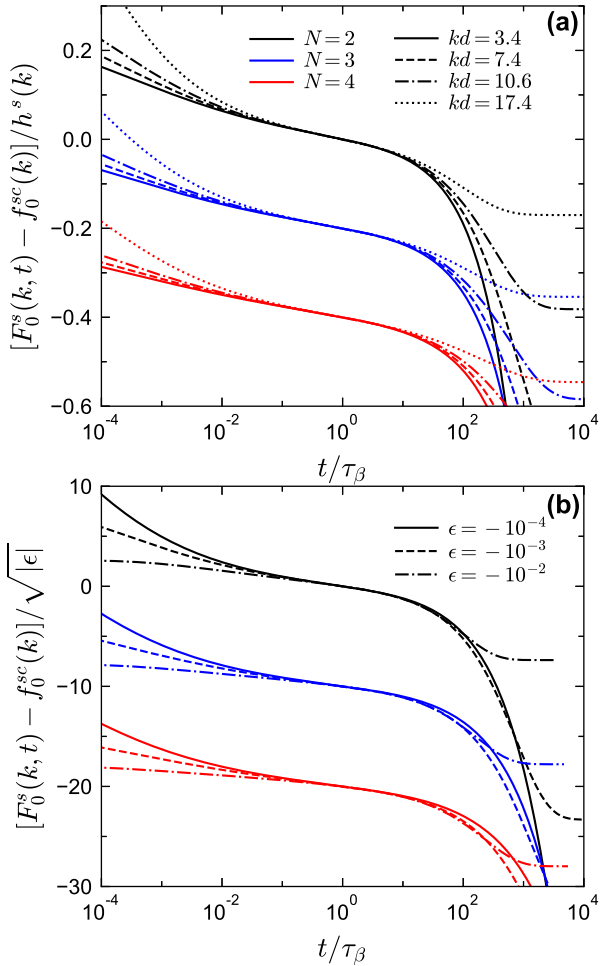}
	\caption{\label{fig:beta_scaling} $\beta$-relaxation scaling laws for different MF closure levels. (a) Scaling with the critical amplitude $h^s(k)$ at $\epsilon=-0.001$. The relative correlation functions are normalized by the corresponding $h(k)$ at four different wavenumbers:  $kd=3.4$ (solid lines), $kd=7.4$ (dashed lines), $kd=10.6$ (dash-dotted lines), and $ kd=17.4$ (dotted lines). For clarity, the lines are shifted vertically by $0.2\times (2-N)$ for every level $N$. (b) Scaling with $\epsilon$ for wavenumber $kd=7.4$. The relative correlation functions are scaled by the corresponding $1/\sqrt{|\epsilon|}$ at three different $\epsilon$ values: $\epsilon=-10^{-4}$ (solid lines), $\epsilon=-10^{-3}$ (dashed lines), and $\epsilon=-10^{-2}$ (dash-dotted lines). For clarity, the lines are shifted vertically by $10 \times (2-N)$ for every level $N$.}
\end{figure}

Next, we test the scaling of $F_0^s(k,t)$ with the wavenumber $k$ [Eq.~(\ref{eq:beta_scaling})] using the critical amplitudes of Fig.~\ref{fig:hsq}.  Figure \ref{fig:beta_scaling}(a) shows the rescaled relative self-intermediate scattering functions $|F_0^s(k,t)-f^{sc}_0(k)|/h^s(k)$ at four different wavenumbers and $\epsilon=-10^{-3}$ for all considered MF closure levels. It can be seen that for a given closure level $N$ and reduced packing fraction $\epsilon$, all curves collapse onto one curve around $t=\tau_{\beta}$. This collapsed curve should then be proportional to $g_-(t/\tau_{\beta})$ and corroborates the $k$-dependent scaling behavior of Eq.~(\ref{eq:beta_scaling}). 

We complete our analysis of the $\beta$-relaxation regime by testing the scaling of $G(t)$ with $\epsilon$ [Eq.~(\ref{eq:beta_scaling_epsilon})]. Figure \ref{fig:beta_scaling}(b) shows the relative self-intermediate scattering functions $|F_0^s(k,t)-f^{sc}_0(k)|$ normalized by $\sqrt{|\epsilon|}$ at wavenumber $kd=7.4$. For each closure level $N$, all curves again collapse onto a single one near $t=\tau_{\beta}$. This demonstrates the scaling of $G(t)$ with the square-root of $|\epsilon|$,  thus numerically confirming the validity of Eq.~(\ref{eq:beta_scaling_epsilon}) within higher-order GMCT.

\begin{figure}[h!]
	\includegraphics{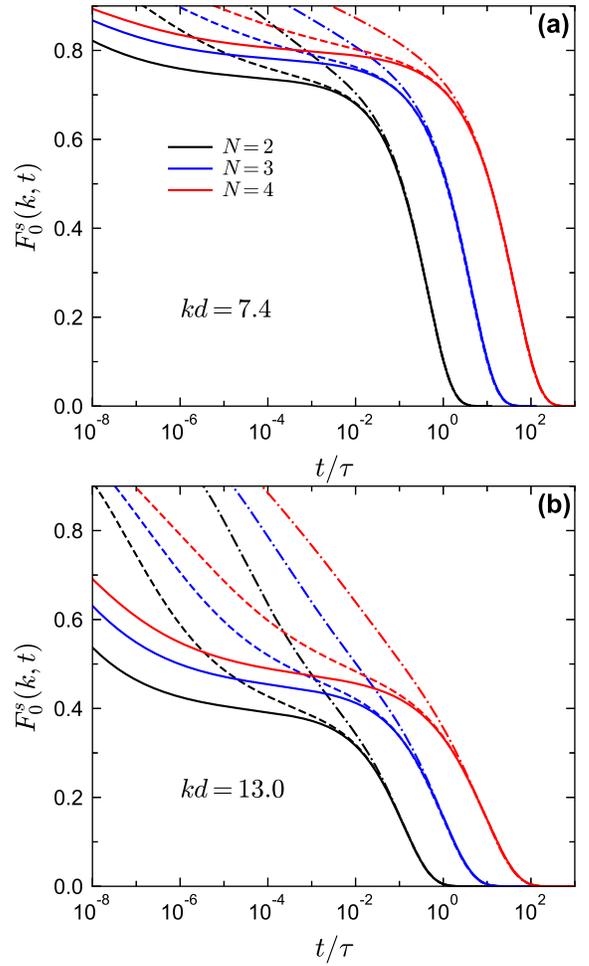}
	\caption{\label{fig:alpha_scaling} $\alpha$-relaxation scaling laws for different MF closure levels. (a) Self-intermediate scattering functions for wavenumber $kd=7.4$ at three different $\epsilon$ values: $\epsilon=-10^{-4}$ (solid lines), $\epsilon=-10^{-3}$ (dashed lines), and $\epsilon=-10^{-2}$ (dash-dotted lines). Different colors correspond to different MF closure levels $N$. For clarity, the lines are shifted horizontally by a factor of $10^{N-2}$. (b) Same as (a) except that $kd=13.0$. For a given closure level $N$ and $\epsilon$, the $\tau$ used here is the same as in (a), defined as $F^s_0(kd=7.4,\tau)=0.1$.}
\end{figure}

Let us now turn to the $\alpha$-relaxation regime. We first seek to test the existence of a time-density (or time-temperature) superposition principle, which is well-established for standard MCT.\cite{gotze2008complex,franosch1997asymptotic,fuchs1998asymptotic} Based on the equivalency of scaling laws for MCT and GMCT on the collective level \cite{luo2020generalized1,luo2020generalized2}, we hypothesize that this superposition principle also holds in the present case. Thus, 
up to order $\sqrt{\epsilon}$, 
the self-GMCT-predicted $\alpha$-relaxation of all $F_0^s(k,t)$ should satisfy the relation 
\begin{equation}
F_0^s(k,t)=\tilde{F}_0^s(k,t/\tau).
\label{eq:alpha_scaling}
\end{equation}
This scaling law absorbs all explicit density (or temperature) dependence into the $\alpha$-relaxation time scale $\tau$ and therefore constitutes a time-density superposition. To substantiate our claim we show the collapse of $F_0^s(k,t)$ onto $\tilde{F}_0^s(k,t/\tau)$ for two different wavenumbers $kd=7.4$ [Fig.~\ref{fig:alpha_scaling}(a)] and $kd=13.0$ [Fig.~\ref{fig:alpha_scaling}(b)] at several reduced packing fractions $\epsilon$. For each closure level $N$, all curves can be seen to collapse, which confirms the scaling law of Eq.~(\ref{eq:alpha_scaling}) in the $\alpha$-relaxation regime. It should, however, be noted that in addition to the different $\tau$ for different $N$, the master function $\tilde{F}_0^s(k,t/\tau)$ also changes with $N$. Thus, each new level in the GMCT hierarchy modifies the quantitative relaxation dynamics.

\begin{figure}[h!]
	\includegraphics{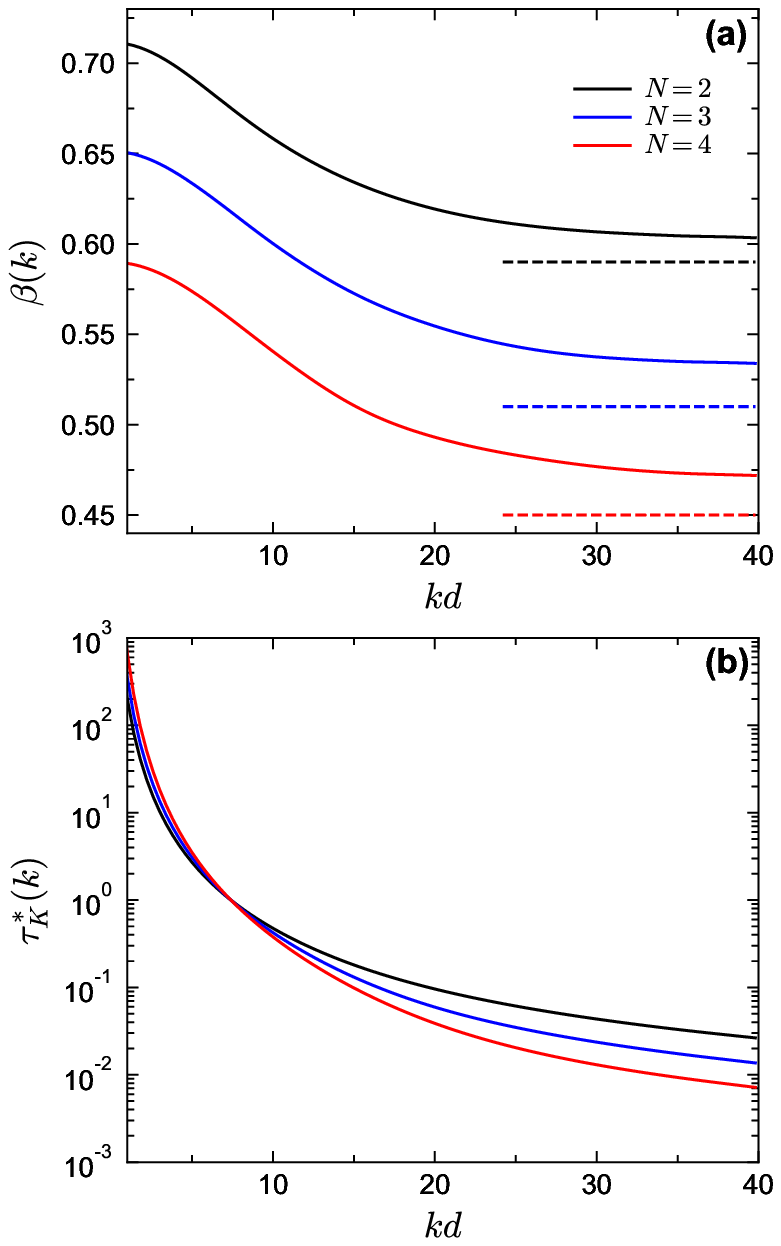}
	\caption{\label{fig:Kohlrausch_fit} Fit parameters for the stretched-exponential Kohlrausch function in the $\alpha$-relaxation regime for different MF closure levels. All curves are obtained by fitting Eq.~(\ref{eq:Kohlrausch}) at $\epsilon=-10^{-3}$. (a) Kohlrausch stretching exponents $\beta(k)$ as a function of wavenumber $k$. The dashed lines are the corresponding parameters $b$ in Table \ref{tab:1}. (b) Rescaled relaxation time $\tau_K^*(k)=\tau_K(k)/\tau_K(k^*=7.4)$. }
\end{figure}
To complete the analysis of the $\alpha$-relaxation regime for liquid states, we  fit our results with the stretched-exponential Kohlrausch function,
\begin{equation}
F_0^s(k,t)=A(k)\exp \left[-\left(\frac{t}{\tau_K(k)}\right)^{\beta (k)}\right],
\label{eq:Kohlrausch}
\end{equation}
for all considered MF closure levels. Figures \ref{fig:Kohlrausch_fit}(a) and \ref{fig:Kohlrausch_fit}(b) show the fit parameters $\beta(k)$ and $\tau^{*}_K(k)=\tau_K(k)/ \tau^{*}_K(k^{*}d=7.4)$ at $\epsilon=-10^{-3}$ (note that we do not show the fitted values of $A(k)$ here since they are almost the same as $f^{sc}_0(k)$ in Fig.~\ref{fig:fc}). We stress that the fit parameters are sensitive to the fitting range (especially for small wavenumbers), hence we have carefully selected the time domains such that at large wavenumbers the fit results are very robust and exhibit only a weak dependence on the fit boundaries. It can be seen that as the closure level $N$ increases, $\beta(k)$ decreases. At the largest wavenumber $k$ used in the numerical GMCT calculations, the value of $\beta(k)$ for each level $N$ is still larger than the corresponding value of $b$. Moreover, the value of $\beta(k)$ is also larger than the $\beta(k)$ obtained from the fit of the collective intermediate scattering functions \cite{luo2020generalized1}. We expect that in the limit of infinite wavenumber, $\beta(k)$ obtained from both the SISF and ISF will converge to the respective value of $b$ for a given $N$ \cite{fuchs1994kohlrausch}. We mention that in simulations of weakly polydisperse hard spheres \cite{weysser2010structural}, the fitted $\beta(k)$ is smaller than the one predicted by MCT (see the lower panel of Fig.~8 in \onlinecite{weysser2010structural}). Interestingly, within GMCT, $\beta(k)$ can be systematically lowered when using higher closure levels $N$, which quantitatively improves its value and allows it to approach the simulation results. 

By contrast, the MCT-predicted values for $\tau^*_K(k)$ are already in reasonable agreement with the simulation data (see the lower panel of Fig.~7 in \onlinecite{weysser2010structural}), and  deviations appear mainly at small wavenumbers where $\tau_K^*(k)$ is overestimated by MCT. Figure \ref{fig:Kohlrausch_fit}(b) demonstrates that by using higher-order GMCT, the deviations become slightly more severe as $N$ increases. However, the qualitative form of the curves remains similar. To test if GMCT improves the prediction quantitatively one would need to compare the unscaled relaxation time $\tau_K(k)$, which has not been provided in \onlinecite{weysser2010structural}.     

\subsubsection{Relaxation dynamics for glass states}
\begin{figure}[h!]
	\includegraphics{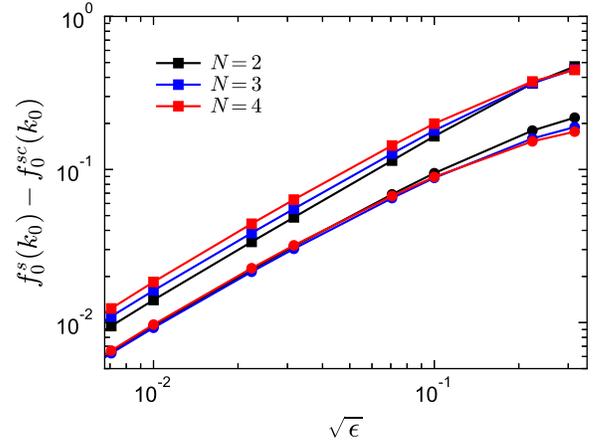}
	\caption{\label{fig:Flong_scaling} Scaling laws for the long-time limit of the self-intermediate scattering function in the glass state [Eq.~(\ref{eq:Flong_scale})] 
	for different GMCT MF closure levels $N$. Circles correspond to wavenumber $k_0d=7.4$ and squares to wavenumber $k_0d=13.0$.}
\end{figure}

In the glass phase, i.e.\ when the packing fraction is larger than the critical packing fraction ($\epsilon>0$), the self-intermediate scattering functions $F_0^s(k,t)$ fail to decay to zero and instead approach some finite positive value $f_0^s(k)$. These long-time limits $f_0^s(k)$ generally increase for increasing $\epsilon$; more specifically, for all  GMCT MF closure levels $N$, they follow (to leading order in $\epsilon$) a scaling law similar to the one derived for MCT \cite{fuchs1998asymptotic}:
\begin{equation}
f_0^s(k)=f_0^{sc}(k)+C\sqrt{\epsilon}h^s(k).
\label{eq:Flong_scale}
\end{equation} 
Here $f_0^{sc}(k)$ and $h^s(k)$ are the same functions introduced in Fig.~\ref{fig:fc} and Fig.~\ref{fig:hsq}, respectively, and $C$ is an $N$-dependent constant. Figure~\ref{fig:Flong_scaling} shows the relative long-time limit of the SISF, $f_0^s(k)-f_0^{sc}(k)$, as a function of $\sqrt{\epsilon}$ at two different wavenumbers. The linear relation at small $\sqrt{\epsilon}$ confirms the scaling with $\epsilon$ in Eq.~(\ref{eq:Flong_scale}).

\begin{figure}[h!]
	\includegraphics{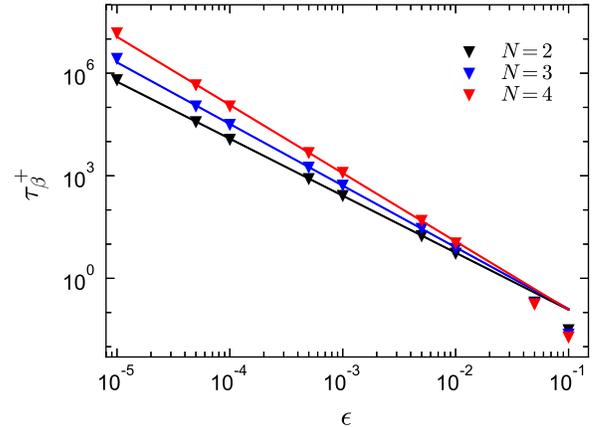}
	\caption{\label{fig:tau_glass} Relaxation times $\tau_{\beta}^+$ of the self-intermediate scattering functions $F^s_0(k,t)$ at $kd = 7.4$ for glass states. Triangles are the numerical $\beta$-relaxation times. The solid lines are the fitted power-law curves $\tau_{\beta}^+=\tau_{\beta 0}^+|\epsilon|^{-1/2a}$. Both $\tau_{\beta 0}^+$ and $a$ are $N$-dependent. }
\end{figure}

Since no second decay step occurs in the glass state, there is only one relevant time scale $\tau_{\beta}^+$ that characterizes the $\beta$-relaxation regime. We have retrieved $\tau_{\beta}^+$, which can be defined as $F^s_0(k,\tau^{+}_{\beta})-f_0^{sc}(k)=1.001(f_0^s(k)-f_0^{sc}(k))$, numerically for each closure level $N$; the results are plotted in Fig.~\ref{fig:tau_glass}. Consistent with the $\beta$-relaxation time $\tau_{\beta}$ obtained in the liquid state, we observe that $\tau_{\beta}^+$ scales with the reduced packing fraction $\epsilon$ as $\tau_{\beta}^+\sim \epsilon^{-1/2a}$. We can thus use $\tau_{\beta}$ to characterize the $\beta$-relaxation regime for both the liquid and glass phases.
The scaling laws of $F^s_0(k,t)$ in the $\beta$-relaxation regime of the glass state are also similar to the ones in the liquid state, i.e.\ Eq.~(\ref{eq:beta_scaling}) and Eq.~(\ref{eq:beta_scaling_epsilon}), except for a change $g_-(t/\tau_{\beta})$ to $g_+(t/\tau_{\beta})$ \cite{fuchs1998asymptotic}. Since these scaling laws are preserved for all closure levels $N\geq2$ in the liquid state, one can expect them to also hold in the glass state, which indeed we have verified numerically (data not shown here).

Summarizing, our results indicate that the scaling laws for the SISF from GMCT in both the liquid and glass state and in both the $\beta$- and $\alpha$-relaxation regime are essentially the same as those predicted by MCT, although the (exponent) parameters $a$, $b$, $\gamma$, $\lambda$, $h^s(k)$, $\tau_K(k)$, and $\beta(k)$ change as the closure level $N$ increases. The values of the main exponents $a$, $b$, $\lambda$, and $\gamma$ are also consistent with those obtained from the ISF for Percus-Yevick hard spheres. Therefore, the remarkably successful scaling laws in MCT are fully preserved in GMCT under generalized mean-field closures. 
We mention that the here discussed scaling laws for $F_0^s(k,t)$ are also applicable to higher-order density correlators $F_n^s(k_0,\{k_i\},t)$, similar to the case of the collective multi-point density correlators \cite{luo2020generalized2}. This is because the mathematical structure of the GMCT equations is essentially identical for all levels of the hierarchy; the vertex only modifies the wavenumber-dependent exponents such as $h^s(k)$. 
 
\subsection {Mean-squared displacement and the Stoke-Einstein relation}
\begin{figure}[h!]
	\includegraphics{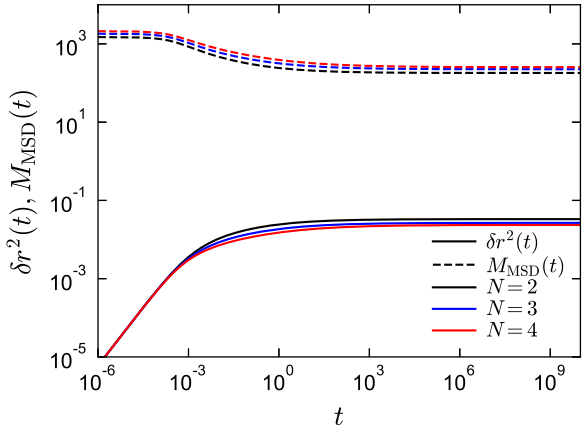}
	\caption{\label{fig:msd_critical} The mean-squared displacement $\delta r^2(t)$ (solid lines) and the associated memory functions $M_{\text{MSD}}(t)$ (dashed lines) at the critical packing fraction $\varphi^c$ for Percus-Yevick hard spheres, obtained from GMCT for different MF closure levels $N$. The calculated localization lengths are $r_s=0.0744,\  0.0667,\ 0.0627$ for $N=2,\ 3,\ 4$, respectively.}
\end{figure}
The MSD can be calculated via Eq.~(\ref{eq:MSD}) and Eq.~(\ref{eq:Mmsd}) once $F_1^s(k_0,k_1,t)$ is known from GMCT. In Fig.~\ref{fig:msd_critical} we show the obtained MSDs $\delta r^2(t)$ (solid lines) at the critical point for all considered closure levels $N$ together with the corresponding memory functions $M_{\text{MSD}}(t)$ (dashed lines). In the long-time limit, $F_1^s(k_0,k_1,t)$ converges to some finite positive value, so that $M_{\text{MSD}}(t)$ also remains positive and $\delta r^2(t)$ becomes constant. We obtain the critical localization length $r_s$ from Eq.~(\ref{eq:rs}) (see the values of $r_s$ in Table \ref{tab:1}) and find that as $N$ increases, $r_s$ decreases, which means that the cage is smaller compared to the prediction from MCT. This is also consistent with the larger $f^{sc}_0(k)$ for higher $N$ shown in Fig.~\ref{fig:fc}.  

\begin{figure}[h!]
	\includegraphics{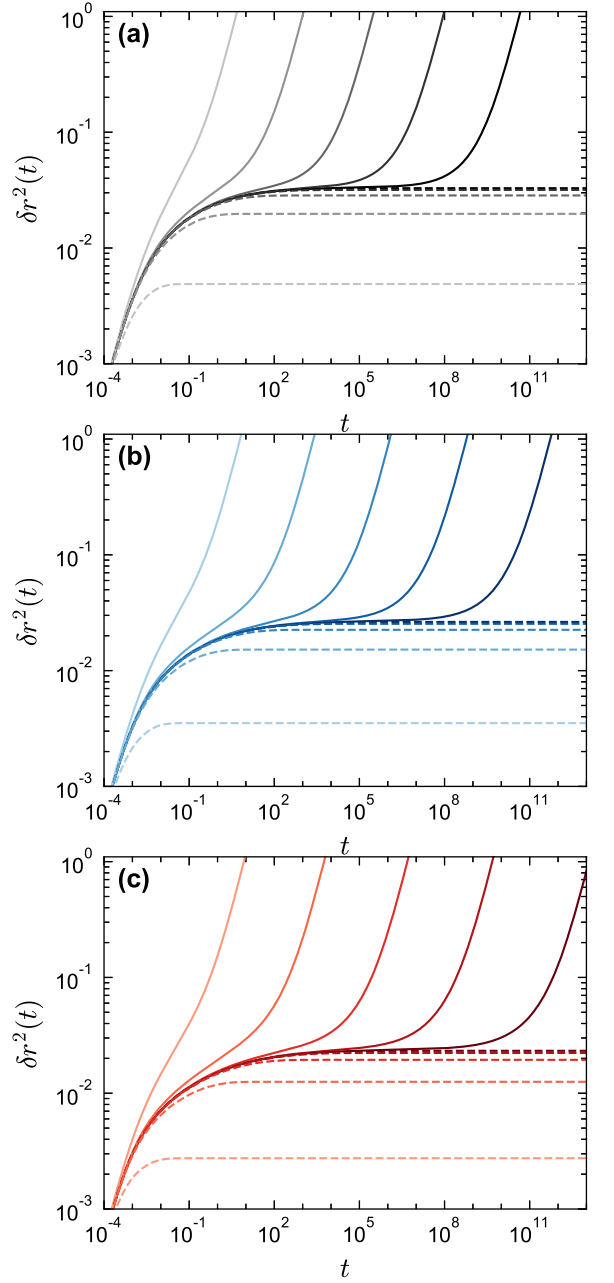}
	\caption{\label{fig:msd} The mean-squared displacement $\delta r^2(t)$ for Percus-Yevick hard spheres obtained from GMCT under  MF closures (a) $N=2$, (b) $N=3$, (c) $N=4$. In all panels, the MSD is shown at $|\epsilon|=10^{-1},\ 10^{-2},\ 10^{-3},\ 10^{-4},\ 10^{-5}$ (from light to dark) for both liquid states (solid lines) and glass states (dashed lines).}
\end{figure}

\begin{figure}[h!]
	\includegraphics{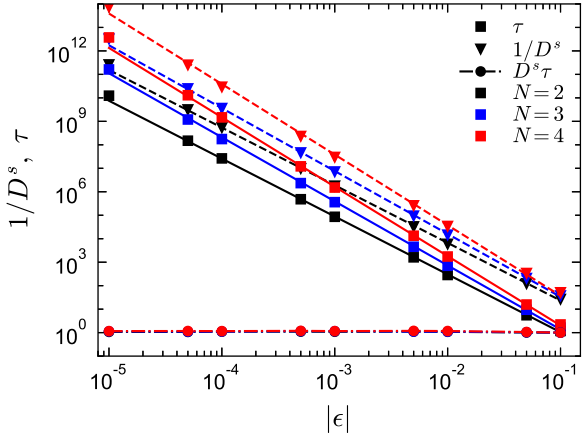}
	\caption{\label{fig:SE_relation} Long-time diffusion coefficient $D^s$ and the Stokes-Einstein relation for different GMCT MF closure levels. Triangles are the inverse of the long-time diffusion coefficient $1/D^s$ and squares are the $\alpha$-relaxation times $\tau$ (same as in Fig.~\ref{fig:tau_scaling}). The solid lines are the fitted power-law curves $\tau=\tau_0|\epsilon|^{-\gamma}$. The dashed lines are the fitted power-law curves $1/D^s=|\epsilon|^{-\gamma}/D^s_0$. The circles are $D^s\tau$ scaled by the value of $D^s\tau$ at $\epsilon=10^{-1}$ and the dash-dotted lines are guides to the eye.}
\end{figure}
Figure \ref{fig:msd} shows the tagged-particle MSDs in both the liquid and glass state at different reduced packing fractions $\epsilon$. It can be seen that for all considered closure levels $N$, the diffusion-localization transition exists: for packing fractions below the critical point, the particle always diffuses in the long-time limit with a diffusion coefficient $D^s$ [as described by Eq.(\ref{eq:Ds})]; for larger packing fractions $\varphi\geq\varphi^c$, the particle instead becomes localized.
In Fig.~\ref{fig:SE_relation} we plot $1/D^s$ as a function of $|\epsilon|$, with  $D^s$ obtained from Fig.~\ref{fig:msd}. We find that the long-time diffusion coefficient follows a power law $D^s\sim |\epsilon|^{\gamma}$ for all GMCT levels, with the same fitted ($N$-dependent) exponent $\gamma$ as  retrieved from the $\alpha$-relaxation time $\tau$ (see Fig.~\ref{fig:tau_scaling}). We also plot the Stokes-Einstein relation, i.e.\ $D^s\tau$, in Fig.~\ref{fig:SE_relation}; it can be seen that the product $D^s\tau$ is constant for all packing fractions close to the critical point. This behavior is strictly satisfied within GMCT for all MF closure levels $N$. Unfortunately, the empirical Stokes-Einstein violation for supercooled liquids is thus not captured by our current GMCT framework, analogous to the case in standard MCT \cite{weysser2010structural, flenner2005relaxation}.

To rationalize the preservation of the SER within GMCT, let us take a closer look at Eq.~(\ref{eq:Mmsd}) and Eq.~(\ref{eq:Ds_M}). Combining both equations gives
	\begin{eqnarray}
	D^s &&=D_0^s \left[1+D_0^s\int_{0}^{\infty}M_{MSD}(u)du\right]^{-1}
	\nonumber\\
	&&=D_0^s \left[1+D_0^s\int_{0}^{\infty}\frac{\rho }{6\pi^2}
	\int_0^{\infty}  p^4 c^2(p)
	F^s_{1}(p,p,u) dpdu\right]^{-1}
		\nonumber\\
	&&=D_0^s \left[1+D_0^s\frac{\rho }{6\pi^2}
	\int_0^{\infty}dp  p^4 c^2(p)
	\int_{0}^{\infty}F^s_{1}(p,p,u) du\right]^{-1}
	\nonumber\\
	&&=D_0^s \bigg[1+D_0^s\frac{\rho }{6\pi^2}
	\int_0^{\infty}dp  p^4 c^2(p)\times
	\nonumber\\
	&&\left(\int_{0}^{\tau_{\alpha^-}}F^s_{1}(p,p,u)+
	\int_{\tau_{\alpha^-}}^{\infty}F^s_{1}(p,p,u) du \right)\bigg]^{-1},
\nonumber\\
\end{eqnarray}
where, to split the time integral, we have introduced $\tau_{\alpha^-}$ as the time scale at which the $\alpha$ relaxation starts. When we are close enough to the critical point this time scale can be taken to satisfy $\tau_{\alpha^-}/\tau\ll 1$. An inspection of both time integrals shows that the first is of the order $O(\tau_{\alpha^-})$ (since in this range $F^s_{1}(p,p,u)\sim 1$), while the second one is at least of the order $O(\tau)$. This implies that we can neglect the first time integration, allowing us to write 
	\begin{eqnarray}
	D^s&&\approx D_0^s \bigg[1+D_0^s\frac{\rho }{6\pi^2}
	\int_0^{\infty}dp  p^4 c^2(p)\int_{\tau_{\alpha^-}}^{\infty}F^s_{1}(p,p,u) du\bigg]^{-1}
\nonumber\\
	&&=D_0^s \left[1+D_0^s\frac{\rho }{6\pi^2}
\int_0^{\infty}dp  p^4 c^2(p)
\int_{\tau_{\alpha^-}}^{\infty}\tilde{F}_1^s(p,p,u/\tau) du\right]^{-1}
\nonumber\\
	&&= D_0^s \left[1+\tau D_0^s\frac{\rho }{6\pi^2}
\int_0^{\infty}dp  p^4 c^2(p)
\int_{\tau_{\alpha^-}/ \tau }^{\infty}\tilde{F}_1^s(p,p,v) dv\right]^{-1}
\nonumber\\
	&&\approx D_0^s \left[1+\tau D_0^s\frac{\rho }{6\pi^2}
\int_0^{\infty}dp  p^4 c^2(p)
\int_{0 }^{\infty}\tilde{F}_1^s(p,p,v) dv\right]^{-1}
\nonumber\\
&&=D_0^s \left[1+\tau D_0^s\frac{\rho }{6\pi^2}
\int_0^{\infty}dp  p^4 c^2(p)
 \hat{\tilde{F}}_1^s(p,p,s=0) \right]^{-1}
 \nonumber\\
&&= D_0^s \left[1+\tau D_0^s M_R\right]^{-1}
  \nonumber\\
 &&\approx \frac{1}{\tau M_R}.
	\end{eqnarray}
Here $M_R$ is the relative memory function $M_R=\frac{\rho }{6\pi^2}
\int_0^{\infty}dp  p^4 c^2(p)
\hat{\tilde{F}}_1^s(p,p,s=0)$ with the Laplace transform $\hat{\tilde{F}}_1^s(p,p,s)=\int_0^{\infty} dt \tilde{F}_1^s(p,p,t) \exp(-st)$ and in the fourth and final step we employ
the approximations $\tau_{\alpha^-}/\tau\ll 1$ and $\tau  M_R\gg 1$ respectively. The only additional relation we have used is the time-density superposition principle $F^s_{1}(p,p,t)=\tilde{F}^s_{1}(p,p,t/\tau)$, which can be derived and tested for all MF closure levels similar to the superposition principle for $F^s_0(k,t)$ in Eq.~(\ref{eq:alpha_scaling}). Therefore, $D^s\tau=1/M_R$ applies to all closure levels $N$ when approaching the critical point (although the precise value of $M_R$ is $N$-dependent). In other words, as long as the time-density superposition of the $\alpha$-relaxation holds for $F^s_{1}(p,p,t)$, the SER cannot be violated. 

The above finding constitutes a paradox for GMCT near the critical point, since both the time-density superposition principle and the Stokes-Einstein violation are experimentally observed, yet they cannot hold simultaneously within the confines of GMCT. Of course, for packing fractions far away from the critical point, the SER can be violated trivially either with a small value of $\tau M_R$ (fast relaxation), or by breaking the time-density superposition principle, or even both. However, the time-density superposition principle and the Stokes-Einstein violation are well-established phenomena close to the glass transition point. This indicates that the framework of GMCT still needs to be qualitatively improved to be able to capture the full phenomenology of glassy dynamics.



\section{Conclusion}
In this work, we have developed a generalized mode-coupling theory for tagged-particle motion. Using the static structure factor $S(k)$ as the only material-dependent input, the theory predicts the  microscopic relaxation dynamics of an arbitrary particle within a dense glass-forming system. The main new equations are the equations of motion for the tagged-particle multi-point density correlation functions [Eqs.~(\ref{eq:GMCTF_sn})-(\ref{eq:Fsclosure})] and the equations for the mean-squared displacement [Eqs.~(\ref{eq:MSD})-(\ref{eq:Mmsd})].

The newly developed theory has been applied to the Percus-Yevick hard sphere system using different mean-field closures. We have comprehensively studied the dynamics of the self-intermediate scattering function near the liquid-glass transition point, demonstrating that the SISF decays in the same manner as its collective counterpart, the ISF \cite{luo2020generalized1,luo2020generalized2}. In particular, for a given packing fraction, increasing the mean-field closure level $N$ yields faster relaxation dynamics, while for a given reduced packing fraction, i.e.\ at a same relative distance to the critical point, the relaxation dynamics instead become slower as more levels are incorporated into the theory. The latter effect is also reflected in the increase of the self-non-ergodicity parameters $f^{sc}_0(k)$ at the critical point as the closure level $N$ increases. Therefore, the fact that increasing $N$ effectively enhances the non-linear feedback mechanism for the collective GMCT \cite{luo2020generalized1} is also correct for the tagged-particle GMCT.

The asymptotic scaling laws of the SISF for both the liquid and glass state, which include the power laws governing the characteristic time scales, the time-wavenumber factorization in the $\beta$-relaxation regime, the time-density (or time-temperature) superposition principle, and the Kohlrausch stretching in the $\alpha$-relaxation regime  are all similar to those of the SISF within standard MCT at all GMCT MF closure levels considered. However, the corresponding  parameters $a$, $b$, $\gamma$, $\lambda$, $h^s(k)$, $\tau_K(k)$, and $\beta(k)$ all explicitly depend on the closure level $N$. The main power-law exponent parameters $a$, $b$, $\lambda$, and $\gamma$ agree with the corresponding values extracted from the predicted ISF within GMCT and have been shown to be closer to the empirical values with increasing $N$. The Kohlrausch stretching exponent $\beta(k)$, which converges towards $b$ in the long-wavenumber limit, has also been shown to reach better agreement with simulations as $N$ increases. The corresponding scaled $\alpha$-relaxation time $\tau^{*}_K(k)$ instead remains qualitatively similar to the MCT result upon increasing the GMCT closure level, although we expect the actual relaxation time $\tau_K(k)$ to be quantitatively improved.   
Interestingly, by contrast, the GMCT-predicted critical amplitude $h^s(k)$ seems to increasingly deviate from the simulation data, with the predicted peak value shifting towards larger wavenumbers. This may be linked to the smaller predicted cage length $r_s$ at higher closure levels. 
Overall, we conclude that the higher-order GMCT framework for tagged-particle motion inherits virtually all qualitative features of standard MCT; however, all numerical predictions for the dynamics change quantitatively upon increasing the closure level, and for most properties studied here we find systematic numerical improvement.  

Unfortunately, our present theory cannot predict the Stokes-Einstein violation for supercooled liquids. In fact, within GMCT, we have proved that the time-density (or time-temperature) superposition principle and the Stokes-Einstein violation are in conflict with each other, and that the current theory is unable to account for both phenomena simultaneously--regardless of the closure level. Therefore, qualitative improvements of the framework, e.g.\ by including the previously neglected off-diagonal terms of the higher-order correlators in the memory function \cite{SimonePhdthesis} or by changing the set of slow variables in the Zwanzig-Mori projection operator \cite{charbonneau2018microscopic}, are needed to adequately capture  Stokes-Einstein violation. More generally, such additional efforts are likely necessary to fully account for the emergence of dynamical heterogeneity, in particular to describe hopping and facilitation effects or activated dynamical processes \cite{charbonneau2013dimensional,bhattacharyya2008facilitation} on a strictly first-principles basis.

\begin{acknowledgments}
We acknowledge the Netherlands Organisation for Scientific Research (NWO) for financial support through a START-UP grant.
\end{acknowledgments}

\section*{AIP Publishing Data Sharing Policy
}
The data that support the findings of this study are available from the corresponding author upon reasonable request.

\bibliography{selfGMCT}

\end{document}